\documentclass[a4paper]{jpconf}

\usepackage{graphics}
\usepackage{adjustbox}
\usepackage{lineno}
\usepackage{subcaption}
\usepackage{placeins}
\usepackage{hyperref}
\usepackage{iopams}

\begin{document}
\title{A First Application of Collaborative Learning In Particle Physics}
\author{Stefano Vergani$^{1,2}$, Attila Bagoly$^2$}
\address{$^1$ University of Cambridge, Cambridge CB3 0HE, United Kingdom}
\address{$^2$ Fetch.AI}
\ead{sv408@hep.phy.cam.ac.uk}
\begin{abstract}
Over the last ten years, the popularity of Machine Learning (ML) has grown exponentially in all scientific fields, including particle physics. The industry has also developed new powerful tools that, imported into academia, could revolutionise research. One recent industry development that has not yet come to the attention of the particle physics community is Collaborative Learning (CL), a framework that allows training the same ML model with different datasets. This work explores the potential of CL, testing the library Colearn with neutrino physics simulation. Colearn, developed by the British Cambridge-based firm Fetch.AI, enables decentralised machine learning tasks. Being a blockchain-mediated CL system, it allows multiple stakeholders to build a shared ML model without needing to rely on a central authority. A generic Liquid Argon Time-Projection Chamber (LArTPC) has been simulated and images produced by fictitious neutrino interactions have been used to produce several datasets. These datasets, called learners, participated successfully in training a Deep Learning (DL) Keras model using blockchain technologies in a decentralised way. This test explores the feasibility of training a single ML model using different simulation datasets coming from different research groups. In this work, we also discuss a framework that instead makes different ML models compete against each other on the same dataset. The final goal is then to train the most performant ML model across the entire scientific community for a given experiment, either using all of the datasets available or selecting the model which performs best among every model developed in the community.
\end{abstract}


\section{Collaborative Learning Framework and Colearn Library}

\begin{figure}[!ht]
\scalebox{1.2}{\includegraphics[width=0.5\textwidth]{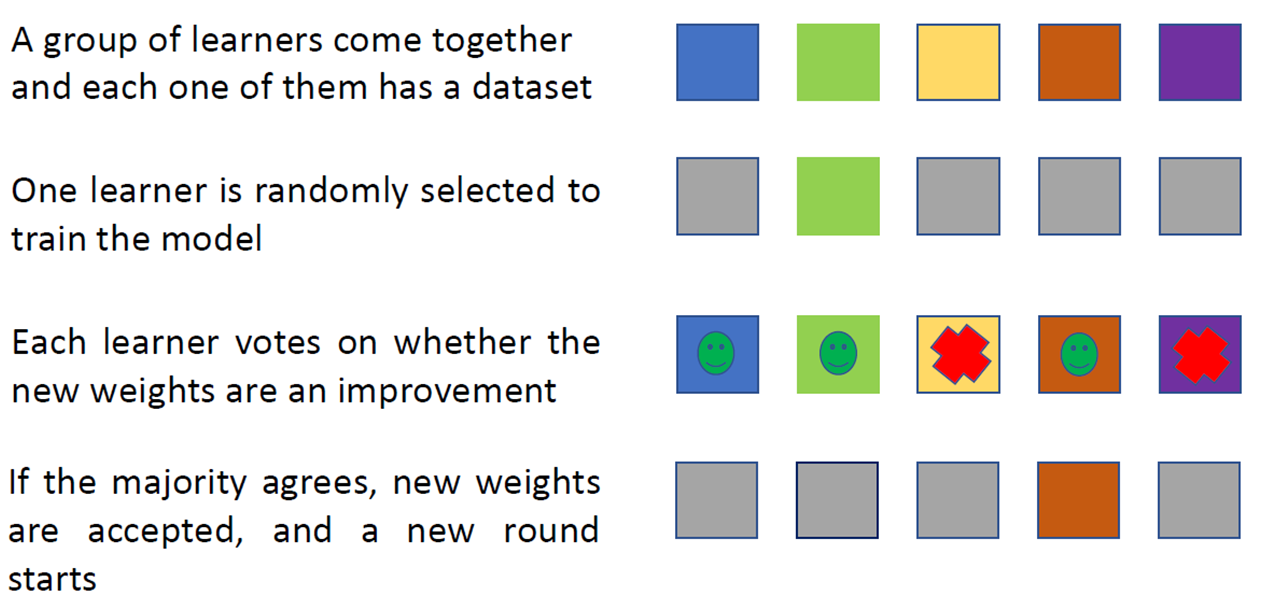}}
\centering
\caption{The Colearn process.}
\label{fig:colearn}
\end{figure}

\begin{figure}[!ht]
\scalebox{1.2}{\includegraphics[width=0.5\textwidth]{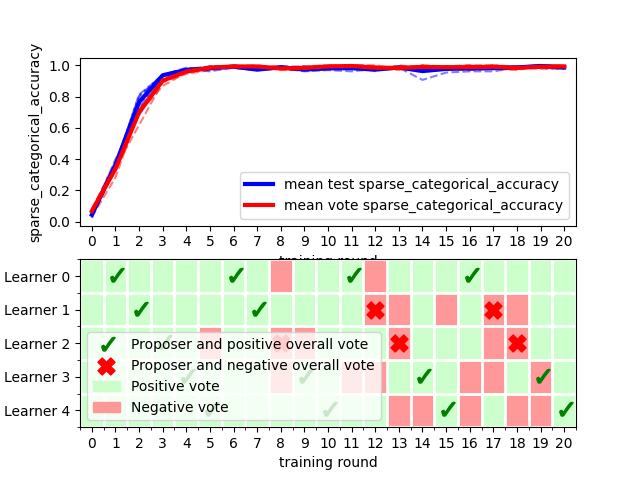}}
\centering
\caption{A Keras model trained using the MNIST data set split into 5 sub data sets. Top: categorical accuracy for each learner (dotted lines) and average (full lines) for test (blue) and vote (red). Below: votes for each training round. Green means the new weights have been approved, red means the new weights have not been approved. When two or more learners approved the weights, the new weights have been overall approved.}
\label{fig:mnist_keras}
\end{figure}

Various methods have been proposed to train the same ML model with separate data sets. For the industry, there has been a strong focus on creating new frameworks to preserve privacy. An example is Federated Learning (FL) \cite{DBLP:conf/aistats/McMahanMRHA17,49232}, which allows the model to be sent directly to the source of data. A typical example could be an ML model used to recognise writing patterns. A company starts with a standard model pre-trained in a certain language and then it sends it to a smartphone. The model will be trained locally with the writing patterns contained in the device and, after the training phase, it will be sent directly back to the warehouse. In this way, the users will never need to send their data directly to the company, only the updated model is shared. Adding noise during training \cite{10.1145/2976749.2978318} will improve privacy even more. A different approach is Collaborative Learning (CL), which is the framework Colearn \cite{colearn_software} is based on. Colearn is a library, created by the company Fetch.AI, that allows privacy-preserving decentralized machine learning tasks using blockchain technologies. This library is particularly interesting because it can be used to simultaneously train the model with several data sets as well as determine the quality of each data set concerning its ability to train the chosen model. The process begins when each stakeholder, that are called learners, bring its data set. This data set will not be disclosed to the other participants, which is a particularly important feature for stakeholders who value privacy. One learner is then randomly selected to train the model, and at the end of the training, the weights are shown to the other learners. They will then test the model with these new weights on their subset and vote on whether the set of weights constitutes an improvement or not. If the majority agrees, the new weights are accepted and a new round starts. For a visual representation of the process, see Figure \ref{fig:colearn}. A first test was done using the MNIST data set \cite{lecun-mnisthandwrittendigit-2010} and breaking it into five sub data sets and training for 20 rounds as it can be seen in Figure \ref{fig:mnist_keras}. Whilst new weights from every data set tend to be accepted during the first rounds, after a while, it can be seen how some sub data sets will be voted more than others. Moreover, the model has been successfully trained using the 5 learners and reaching an accuracy above 95\%.

\section{Particle Physics and Collaborative Learning} \label{sec:pp_and_cl}

The field of particle physics has seen an exponential usage of ML technologies over the last years \cite{calafiura2022, PRESSCUT-H-2018-405}. It is customary for different groups or experiments to produce their simulated data set to train ML models or use data from real experiments. This work intends to explore the capabilities of CL with particle physics data sets. One possible usage could be merging different simulated or real data sets to train the same ML models. This could be useful for example for training an ML model to look for several low-interacting particles or rare events, where each group can contribute bringing a sub data set focused only on a particular particle or events. Another option could be instead of making several data sets compete against each other to determine which data set is the most promising one. Given that Colearn is decentralised, this could a particularly useful feature to ensure impartiality among different research groups. The last option, that will not be discussed in this paper, is to make different ML models compete against each other with the same data set. This is in theory possible but at the moment there is no such implementation on the market.

\section{The Neutrino Data Set} \label{sec:dataset}

\begin{figure}[!ht]
\scalebox{2}{\includegraphics[width=0.5\textwidth]{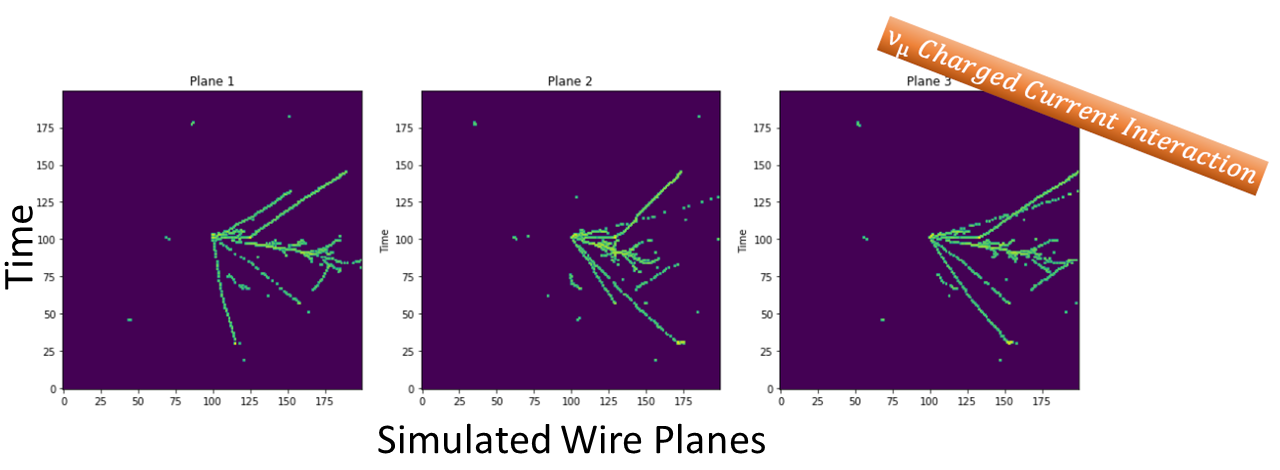}}
\centering
\caption{One example of a muon NCC from the neutrino data set. The three images mimic the three wire plane views as seen in a LArTPC experiment.}
\label{fig:wires}
\end{figure}

For this proof of concept, a neutrino data set has been created consisting of 10000 training and 2000 validation pictures. Each picture contains three sub-images that mimic the three wire-planes of a typical Liquid Argon Time-Projection Chamber (LArTPC) such as a MicroBooNE \cite{Acciarri_2017, duffy2022a} or ProtoDUNE Single Phase (SP) \cite{DUNE:2020cqd,DUNE:2021hwx}. Each picture is $200\times200$ pixels and represent either a Neutrino Neutral Current (NNC), a Muon Neutrino Charged Current (NCC), or an electron NCC. For an example of the data set, see Figure \ref{fig:wires}. The neutrino interactions were created using GENIE \cite{genie} software with a neutrino flux as a function of energy following a Gaussian distribution with a mean of 2.5\ GeV and width 0.5\ GeV. The final-state particles from the neutrino interactions are passed into the GEANT4~\cite{geant} simulation, being tracked through a monolithic liquid argon volume.

\section{Machine Learning Models} \label{sec:keras}

\begin{figure}[!ht]
\scalebox{1.7}{\includegraphics[width=0.5\textwidth]{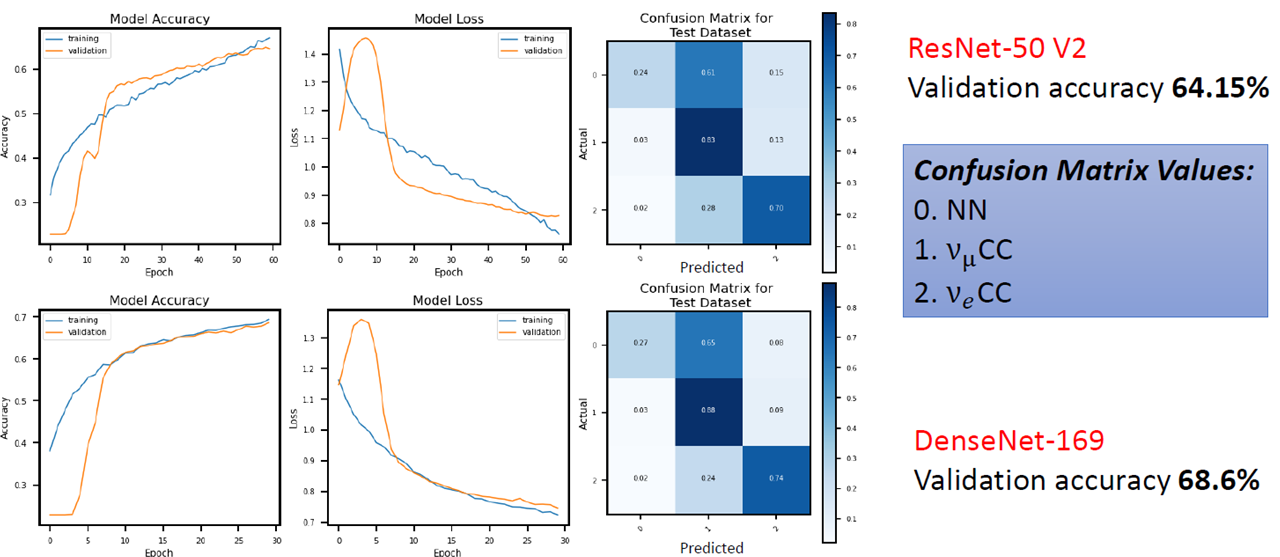}}
\centering
\caption{The Keras model ResNet-50 V2 (top) and DenseNet-169 (bottom) trained with the entire neutrino data set. On the right-hand side is the confusion matrix for each one of them. The accuracy curve suggests that even higher validation accuracy could have been achieved by extended training. These performances represent only a term of comparison to benchmark the proof of concept.}
\label{fig:models}
\end{figure}

For this experiment, several available Keras \cite{chollet2015keras} models have been tested with the neutrino data set mentioned in Section \ref{sec:dataset}. The two best performing have been ResNet-50 V2 and DensNet-169, with an accuracy of 64.15\% and 68.6\% as shown in Figure \ref{fig:models}. Their performance has been used as a benchmark. 

\section{Collaborative Learning with Neutrino Data Set}

\begin{figure}
\centering
\begin{subfigure}[a]{0.55\textwidth}
    \scalebox{2.4}{\includegraphics[width=0.5\textwidth]{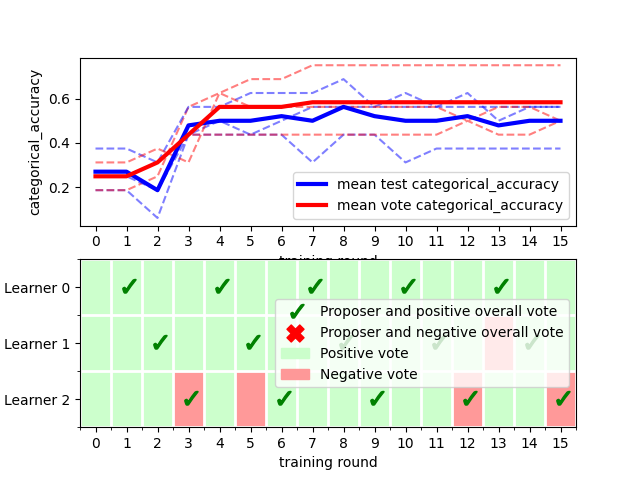}}
    \centering
   \caption{}
   \label{fig:resnet_results} 
\end{subfigure}

\begin{subfigure}[b]{0.55\textwidth}
   \scalebox{2.4}{\includegraphics[width=0.5\textwidth]{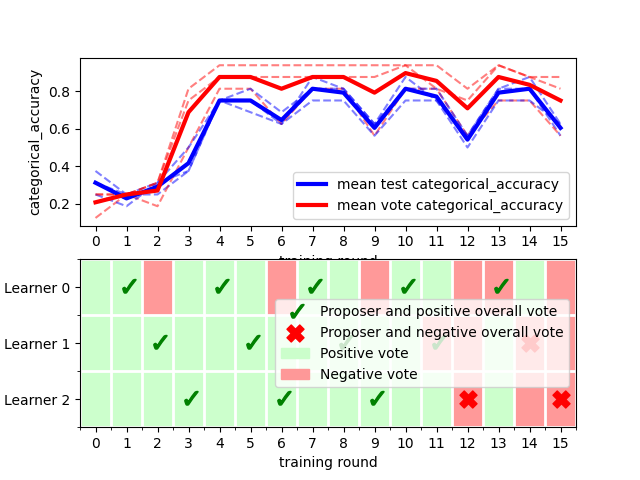}}
   \centering
   \caption{}
   \label{fig:densent_results}
\end{subfigure}

\caption{Test results with Keras (a) ResNet-50 V2 and (b) DenseNet-169. Top: categorical accuracy for each learner (dotted lines) and average (full lines) for test (blue) and vote (red). Below: votes for each training round. Green means the new weights have been approved, red means the new weights have not been approved. When two or more learners approved the weights, the new weights have been overall approved.}
\end{figure}

To simulate the usage of separate data sets, the neutrino data set has been randomly split into three separate subsets. These three sub sets have been used as three learners. Both the models explained in Section \ref{sec:keras} have been trained using the Colearn library for 15 rounds. Results are shown in Figures \ref{fig:resnet_results} and \ref{fig:densent_results}.



\section{Results and Future Applications} \label{sec:results}

This proof-of-concept has been successful and it showed that the Colearn library can be used for particle physics's data sets. Different behaviours are observed between ResNet-50 V2 and DenseNet-169. The former tends to be more stable but with lower efficiency (roughly 50\% the mean test accuracy), the latter presents an unstable but higher efficiency, with peaks around 70\%. In both cases, a certain learner (learner 2 for ResNet-50 V2 and learner 0 for DenseNet-169) was particularly underperforming in terms of having its set of weights approved. This is due to random statistical differences in the sub-samples, but the fact that the framework is correctly able to identify those differences shows its great potentialities. These results confirm that the ideas expressed in Section \ref{sec:pp_and_cl} can now be developed into a proper and independent framework based on Colearn. This could be made to find which simulated data set underperform in a specific task since it would be the learner that most often does not bring accepted weights. It could also be used pretty much straight out of the box to train the model with separate data sets. Further studies on the oscillating behaviour of some models should be made to determine whether this is a feature that can be used. The next step would most importantly be to try a similar experiment with completely different data sets coming from different sources. This would highlight the possible limits of this technology or its further potentialities.\\
\FloatBarrier
\section*{References}
\bibliographystyle{JHEP}
\bibliography{acat}

\end{document}